# Simultaneous nanoscale imaging of local conductivity and chemical potential in a quantum Hall isospin ferromagnet

Jiawei Hu[1,2,#], Shiyu Zhu[1,2,#]✉, Bohao Li[3,#], Yunhao Wang[1,2,#], Shuigang Xu[4], Zhihai Cheng[5], Chengmin Shen[1,2], Andre K. Geim[6], Fengcheng Wu[3]✉, Hong-Jun Gao[1,2]✉

[1]Beijing National Laboratory for Condensed Matter Physics, Institute of Physics, Chinese Academy of Sciences, Beijing 100190, China

[2]School of Physical Sciences, University of Chinese Academy of Sciences, Beijing 100049, China

[3]School of Physics and Technology, Wuhan University, Wuhan 430072, China

[4]Key Laboratory for Quantum Materials of Zhejiang Province, Department of Physics, School of Science, Westlake University, Hangzhou 310030, China

[5]Key Laboratory of Quantum State Construction and Manipulation (Ministry of Education), School of Physics, Renmin University of China, Beijing 100872, China

[6]Department of Physics and Astronomy, University of Manchester, Manchester, UK

[#]These authors contributed equally: Jiawei Hu, Shiyu Zhu, Bohao Li, and Yunhao Wang.

✉ Corresponding author, E -mail: syzhu@iphy.ac.cn; wufcheng@whu.edu.cn; hjgao@iphy.ac.cn.

## Abstract

Quantum Hall isospin ferromagnetism in multilayer graphene offers a versatile playground for exploring flat band correlated physics, driven by the intricate coupling of spin, valley, orbital, and layer degrees of freedom. However, a nanoscale probe capable of simultaneously mapping local conductivity and chemical potential in these exotic phases has yet to be realized. Here, we introduce scanning conductivity and chemical potential microscopy (SCCM), a technique integrating scanning microwave impedance microscopy and Kelvin probe force microscopy. We demonstrate SCCM by probing the quantum Hall states and many-body Landau level energy spectrum in bilayer graphene. Applied to marginally twisted double bilayer graphene, SCCM then reveals a cascade of quantum Hall isospin ferromagnetic states with unexpected re-emergence behaviors. Significantly, experimental many-body Landau level energy spectrum further uncovers the intricate connections of these complex phenomena to inter-subband Landau level crossings and Landau level single-particle wavefunctions. These insights enable the construction of a comprehensive quantum Hall phase diagram. Our results demonstrate SCCM's capability in decoding complex quantum phenomena, establishing it as a versatile nanoscale probe for electron correlation and topology.

## Introduction

Graphene Landau levels (LLs), characterized by quenched kinetic energy and enhanced electron-electron interactions, provide an exceptional platform for exploring flat bands and strongly correlated physics. At integer LL fillings, the fourfold spin-valley degeneracy of Dirac electrons is lifted by exchange interactions, giving rise to an SU(4) quantum Hall isospin ferromagnetic (QHIFM) phase, where the isospin is either polarized or forms a coherent superposition of spin and valley degrees of freedom[1]. QHIFM phases have been extensively investigated in various systems, including graphene[2–6] and its moiré superlattices[7,8], revealing insights into local isospin textures, thermodynamic properties, and transport characteristics. Therefore, investigating the broken-symmetry orders and QHIFM phases in two-dimensional (2D) systems are crucial for advancing the understanding of strongly correlated physics.

QHIFM phases in graphene have been characterized by various scanning probe microscope (SPM) techniques, including scanning tunnelling microscope[8,9], scanning single-electron transistor[10–12], scanning microwave impedance microscopy (sMIM)[13], and Kelvin probe force microscopy (KPFM)[14]. Among these, sMIM detects local conductivity and topology of quantum states at nanoscale by varying gate voltage and magnetic field. This technique has been applied to investigate topologically non-trivial edge states[13,15–19] and correlated insulating states[20–24] in 2D materials, demonstrating its potential for uncovering emergent electronic states in strong correlated systems. Utilizing top gate graphene as a sensor layer, sMIM has recently been employed to probe local chemical potential and thermodynamic energy gaps in correlated states[25,26]. Additionally, KPFM measures the local contact potential difference (CPD) between the tip and sample, detecting the variations in chemical potential and band-filling condition as carrier density changes in the sample, providing complementary insights into quantum Hall (QH) states[14]. Notably, the integration of local conductivity and chemical potential measurements would provide unique insights into the energy gaps, electron interactions, and band topology characteristic of QHIFM systems[27–29]. This highlights the need for advanced local probe techniques that can simultaneously capture both parameters without mutual interference, facilitating a deeper understanding of strongly correlated phenomena in complex materials.

In this study, we realize local conductivity and chemical potential measurement in parallel by developing a technique, SCCM. We exhibit the functionality of SCCM by investigating QH physics both in bilayer graphene (BG) and marginally twisted double bilayer graphene (TDBG, twist angle < 0.2°). In BG, we identify a sequence of QH states and the LL energy spectrum, verifying SCCM capability of probing both local conductivity and chemical potential. In TDBG, we observe robust QHIFM phases at LL filling factor $\nu = +4$ to -16. We show that the many-body LL energy spectrum is generated by two sets of subbands, with the crossing between them causing the re-emergence behaviors, in excellent consistence with theoretical calculation. We therefore establish concrete connections between the LL wavefunctions, LL energy spectrum and the resulting QHIFM phase diagram, highlighting SCCM as a powerful probe for unravelling the complex physics and topological properties of quantum material.

## SCCM technique

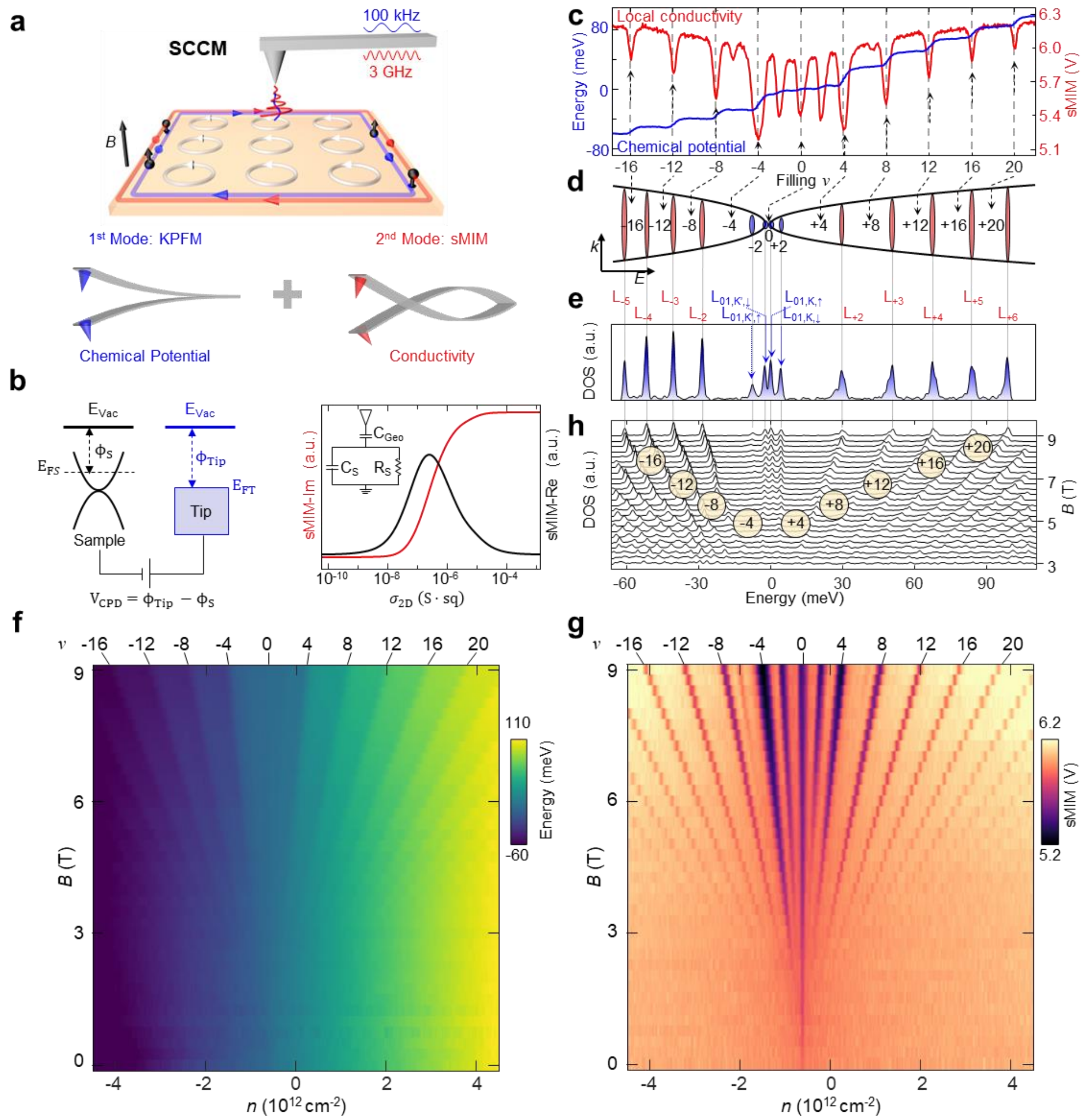

**Fig. 1 | The SCCM technique for measuring local conductivity and chemical potential simultaneously. a**, Schematic of SCCM technique. **b**, Left: principle of KPFM. The contact potential difference $V_{CPD}$ between tip and sample can induce electrostatic force on cantilever. The chemical potential of sample can be accessed by compensating this capacitive force. Right: Principle of sMIM (right). Simulated sMIM response curves as a function of the 2D conductivity of the sample based on the tip-sample effective circuit model (right, inset). **c**, The local conductivity (left axis) and chemical potential (right axis) of QH states of BG as a function of LL filling number $\nu$ at $T = 100$ mK and $B = 9$ T. **d**, Schematic of BG massive Dirac fermion LL energy spectrum. The Chern numbers are labelled in LL gaps. **e**, Thermodynamic DOS of BG extracted from the chemical potential curve in c. **f**, **g**, $n$-$B$ maps of chemical potential (f) and local conductivity (g) in BG region. **h**, BG LL energy spectrum as a function of $B$ deduced from f. Each LL manifests nearly linear dispersion with $B$, consistent with equation (1) in main text.

SCCM integrates sMIM and KPFM to enable simultaneous nanoscale mapping of local conductivity and chemical potential (Fig. 1a and Supplementary Fig. 1). In both methods, a metallic atomic force microscopy (AFM) probe approaches the sample surface with applied AC voltage bias: 3 GHz for sMIM and typically a few kHz frequencies for KPFM. Crosstalk between the two SPM techniques is eliminated by employing different mechanical vibration eigenmodes of the cantilever and corresponding lock-in frequencies for signal demodulation in each channel. In sMIM, the applied microwave bias drives the mobile carriers towards and away from the tip. The imaginary component of the reflected sMIM signal (sMIM-Im) characterizes the sample's ability to screen the electric field. Typically, sMIM-Im exhibits an approximately linear correlation with the logarithm of the sample conductivity in the high sensitivity window (Fig. 1b Right). In KPFM, the oscillating ~kHz ac bias induces an electrostatic force on the cantilever by the capacitive coupling between tip and sample, providing visualization of sample's work function and chemical potential (Fig. 1b Left). During SCCM measurement, the CPD between 2D sample and metallic tip is actively compensated by KPFM feedback, effectively eliminating tip-gating effects.

We demonstrate SCCM through single point gate voltage dependent SCCM spectroscopy in BG region at out-of-plane magnetic field $B$ = 9 T (Fig. 1c). The BG LL energy is approximately given by,

$$E_N = \pm\hbar\omega_C\sqrt{N(N-1)}, \qquad N = 0, 1, 2, \ldots \quad (1)$$

where $\omega_C = \frac{eB}{m^*}$ is the cyclotron frequency, $m^*$ is the effective mass of the quasiparticles and $N$ is the LL index. Here LL with a given $N$ index has four flavors including two spins and two valleys. The zero energy LL has an additional orbital degeneracy between $N$ = 0 and $N$ = 1, giving rise to an eight-fold degenerated zero energy state. The tip, fixed above a specified point on the graphene surface, collects local conductivity and chemical potential during back gate voltage ($V_g$) ramping. The LL filling number $\nu$ is given by $\nu = n/(B/\phi_0)$, where $n$ is carrier density and $\phi_0$ is magnetic flux quantum. At filling $\nu = \pm 4N$, BG chemical potential is tuned to the single-particle energy gap between LLs, thus an incompressible bulk state is formed. A sequence of QH states manifest as conductivity dips and chemical potential jumps at $\nu = \pm 4N$, consistent with Dirac-like quadratic energy spectrum of BG (Fig. 1d). At other integer filling $\nu \neq \pm 4N$, the Fermi level is constrained within the LL flat band, rendering the sample compressible. Therefore, chemical potential plateaus and high conductivity emerge concurrently owing to the compressible state. Moreover, the eight-fold degeneracy in $N$ = 0 and $N$ = 1 LLs is partially lifted, giving rise to a pronounced insulating feature in sMIM channel at $\nu$ = 0 and ±2. Here the zeroth LL splits to four spin-valley polarized subbands, with two-fold orbit degeneracy ($N$ = 0 and $N$ = 1) preserved[4,30]. We extract the thermodynamic density of states (DOS) from the chemical potential curve by gaussian kernel density estimation (Fig. 1e, see "Methods"), wherein the LLs manifest as a series of peaks situated at discrete energies. Our result at 100 mK unveils a discernible correlation between local conductivity and chemical potential in QH physics.

In SCCM, the *n-B* mapping is obtained by systematically varying carrier density (n) and magnetic field (B). Such mapping of chemical potential (Fig. 1f) and local conductivity (Fig. 1g) reveal consistent features of Landau fan diagram, resolving the QH states in both channels. The thermodynamic DOS curves (extracted from Fig. 1f) depict the many-body LL energy spectrum (Fig. 1h). The LLs at $\nu = \pm 4N$ exhibit linear dispersion with $B$ as expected from equation (1). The Chern number of each gap is determined by tracking the chemical potential evolution with $B$ at each integer filling. The LL energies at various $B$ collapse onto a universal curve when plotted as a function of $\pm\sqrt{\mathrm{n}(\mathrm{n}-1)}B$ (Supplementary Fig. 2a), enabling the deduction of the effective masses of electron and hole as $m_e^* = 0.052\, m_e$ and $m_h^* = 0.069\, m_e$. We further extract the parabolic energy-momentum dispersion of BG through the semi-classical Onsager relation[31] $k_{\mathrm{n}} = \sqrt{2e\mathrm{n}B/\hbar}$ (Supplementary Fig. 2b). Our experimental effective band dispersion shows excellent consistence with previous results[32,33]. Consequently, we establish SCCM as a versatile sensor for local conductivity, chemical potential and topological energy spectrum.

## QH states in marginally TDBG

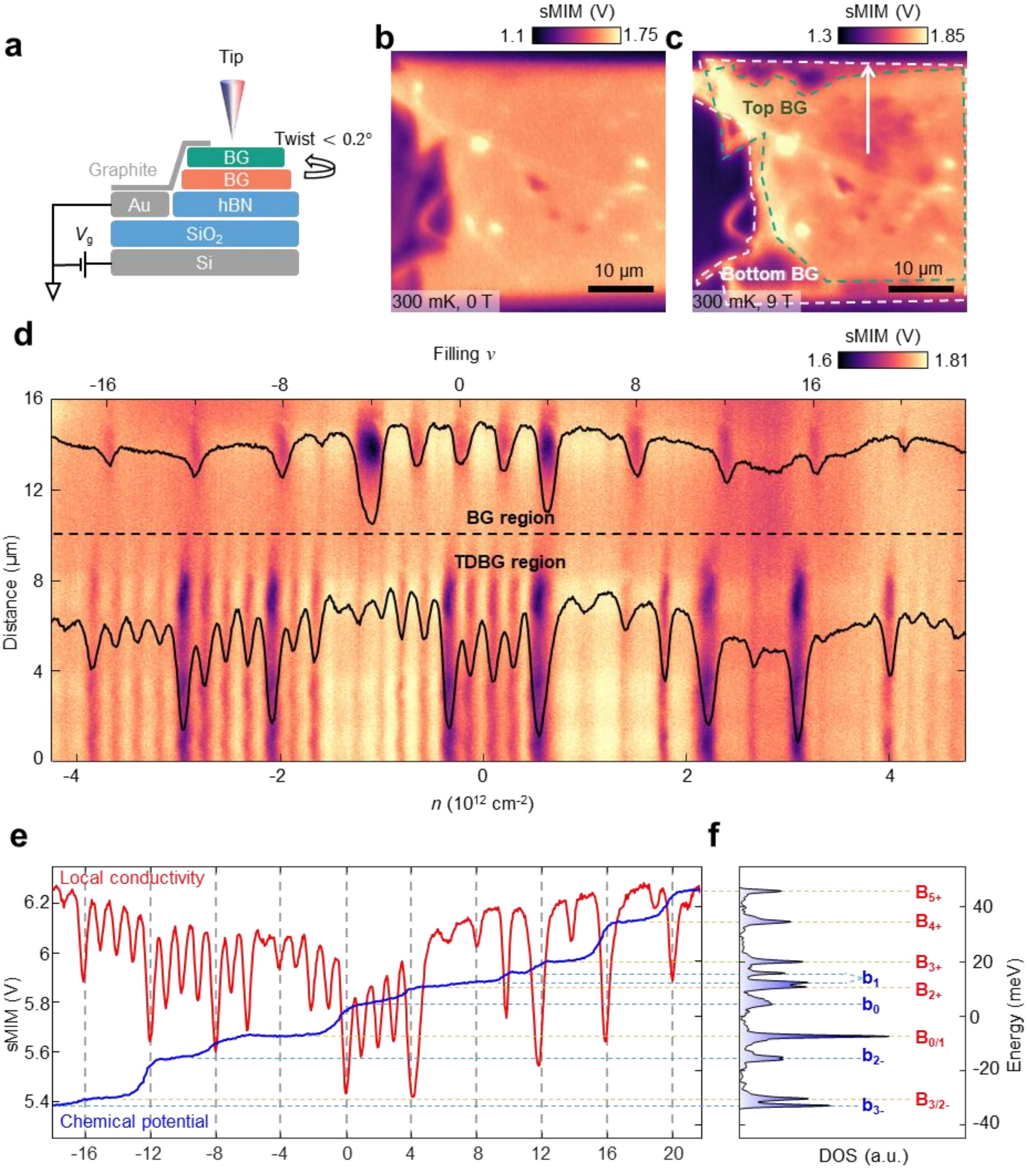

**Fig. 2 | Fully broken-symmetry QHIFM states in TDBG. a,** Schematic of the TDBG device structure. $V_g$ is the gate voltage. **b**, **c**, Large area sMIM mapping of TDBG at $B$ = 0 T (**b**) and 9 T (**c**) and $T$ = 300 mK. The scale bars correspond to 10 μm. **d**, sMIM line scans at $B$ = 9 T along the white arrow indicated in **c**. Repeated line scans (y-axis) are performed while the gate voltage is varied (x-axis). A series of QH states is revealed in both the TDBG and BG regions. Black curves show spatial averaged local conductivity across TDBG and BG region**. e,** The local conductivity (left axis) and chemical potential (right axis) of TDBG as a function of LL filling number $\nu$ at $T$ = 100 mK and $B$ = 9 T. All integer QH states at filling factor from +4 to -16 are clearly resolved in local conductivity, indicating fully broken-symmetry QHIFM states in TDBG. Chemical potential (blue curve) shows a sequence of plateaus, corresponding to LLs indicated in **f**. **f**, Thermodynamic DOS of TDBG extracted from the chemical potential curve in **e**. Each 4-fold LL is denoted by its subband (B or b) and LL index n (n+ for electron-like and n− for hole-like LLs), while the b1 LL is further split into two 2-fold subbands (see Supplementary Fig. 9 for details).

We now turn to TDBG measurement. It has been shown that TDBG with twist angle $\theta$ near 1° hosts a series of novel correlated physics such as spin-polarized correlated insulating phase[34–37], moiré nematic phase[38], and superconductivity[39]. In our study, we fabricate marginally TDBG (twist angle < 0.2°) device on hexagonal boron nitride (hBN) substrate with doped Silicon back gate (Fig. 2a, see "Methods"). Notably, TDBG with hBN placed on a single side has enhanced Coulomb interaction relative to the hBN-encapsulated devices. The local conductivity of the TDBG device is visualized at $B$ = 0 T (Fig. 2b) and 9 T (Fig. 2c). At $B$ = 0 T, the entire graphene device at charge neutrality point is conductive compared to the insulting hBN substrate. In contrast, the BG region exhibits a clear insulating interior and a conductive edge at $B$ = 9 T and $\nu$ = +4, as a hallmark of QH state. Within the TDBG region, multiple insulating blocks emerge due to the formation of incompressible QH states confined by structural wrinkles (Supplementary Figs. 3 and 4)[24]. Zooming into these individual blocks with room temperature sMIM-Re reveals varying domain distributions, which points to slightly different local twist angles near zero. Specifically, some blocks exhibit a mixed domain state of ABAB and ABCA stacking with varying moiré superlattices ranging from 100 nm to 500 nm (Supplementary Fig. 3c–f), with the ABAB domains consistently exhibiting a convexly expanded morphology due to their enhanced stability and lower energy relative to the ABCA configuration. In contrast, extensive regions within other confined blocks completely lack moiré superlattices, having fully relaxed into a uniform ABAB stacking phase (Supplementary Fig. 3g–n). We therefore conclude that the ABAB stacking phase robustly dominates the overall device area, regardless of the local presence of a moiré superlattice.

The QH states in TDBG are visualized through gate dependent sMIM line scanning at $T$ = 100 mK and $B$ = 9 T (Fig. 2d). Both the TDBG and adjacent BG region exhibit a cascade of insulating quantum Hall states. However, the TDBG region displays a notably denser sequence, indicating spin and/or valley symmetry breaking at more filling factors. The QH states maintain a consistent sequence throughout the TDBG region, though the intensity and carrier density fluctuate due to local impurities or wrinkles. We next perform single point gate dependent SCCM spectroscopy in the TDBG region (Fig. 2e). At a base temperature of $T$ = 100 mK and $B$ = 9 T, QHIFM states manifest as clear insulating features at each integer LL filling from $\nu$ = +4 to -16. At LL quarter filling $\nu = 4N \pm 1$, Coulomb interaction lifts the degenerate spin-valley SU(4) isospin, leading to a spin and valley polarized QH ground state which minimizes the exchange energy[3]. At half-filling of the LL ($\nu = 4N \pm 2$), Pauli exclusion hinders simultaneous full polarization of both spin and valley. This creates intricate competition between spin, valley, and/or orbital polarizing tendencies, ultimately determining the resulting ground state. Possible scenario includes the charge density wave state, the spin ferromagnet state, the canted antiferromagnet state and the intervalley coherent state with Kekulé reconstruction[3,4,8,9,40]. We can not determine the exact states from our experiments. In the chemical potential channel, we observe distinct jumps at integer fillings, corresponding to incompressible QH states with relatively large energy gaps. The thermodynamic DOS (Fig. 2f), extracted from the experimental chemical potential curve, reveals a sequence of LL peaks. Determining the precise physical origin of these LLs, however, would require additional information on their dispersion with magnetic field.

## Many-body LL energy spectrum of QHIFM states

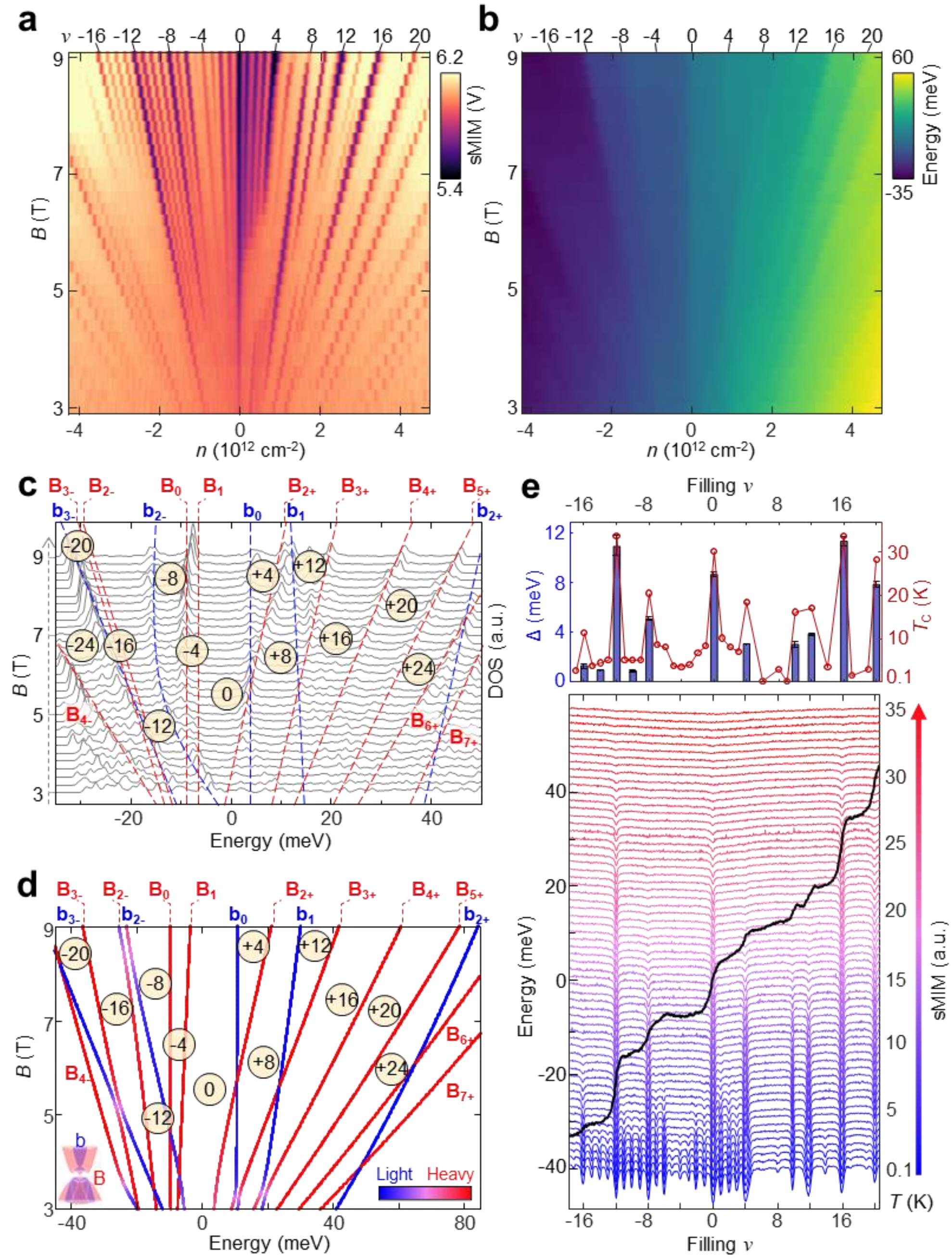

**Fig. 3 | Many-body LL energy spectrum of TDBG. a**, **b**, *n*-*B* map of local conductivity (**a**) and chemical potential (**b**) at $T = 100$ mK in TDBG region. Re-emergence behavior is observed at integer filling factor $\nu = +4, +12, +16$ in **b**. **c**, LL energy spectrum extracted from **b**. Each LL is denoted by its subband (B or b) and LL index n (n+ for electron-like and n− for hole-like LLs). The zeroth LLs are labeled as $b/B_{0/1}$. **d**, Calculated LL spectrum of tetralayer graphene, in good consistent with experimental LL energy spectrum (**c**). Inset: calculated 3D band structure of tetralayer graphene. **e**, Lower: Temperature dependence of the local conductivity as a function filling factor $\nu$ at $B = 9$ T, ranging from 100 mK to 35.6 K. Black curve shows corresponding chemical potential at $T = 100$ mK. Upper: QH states energy gap Δ (blue bar) estimated from chemical potential measurement and their critical temperature $T_C$ (red circle) extracted from temperature dependent measurement. Error bars for Δ represent the uncertainty bounds derived from the transition width of chemical potential curve (see Supplementary Fig. 7 for detailed energy gap extraction procedures).

To further elucidate the symmetry-broken nature of TDBG, we perform $n$-$B$ mapping of SCCM on TDBG (Figs. 3a and 3b). Remarkably, we observe complete lifting of spin and valley degeneracy $\nu$ = +4 to -16 at $B > 5$ T, indicating robust quantum Hall isospin ferromagnetism driven by strong electron-electron interaction (Fig. 3a). These QHIFM states form complex sequences, characterized by the disappearance and re-emergence of certain states as a function of $B$ field, notably at ν = +4, +12, and +16. The chemical potential measurements also reveal a Landau fan-like feature (Fig. 3b), though with lower resolution compared to sMIM data. Notably, the information obtained from sMIM is consistent with that from KPFM, yet sMIM resolves more QH states, underscoring its superior sensitivity to local conductivity variations.

To gain deeper insight into the QHIFM ground states and re-emergence behaviors, we extract the many-body LL energy spectrum (Fig. 3c) of TDBG from the chemical potential Landau fan (Fig. 3b). The Chern number associated with each gap is carefully determined by tracking the correspondence between the simultaneously measured chemical potential and conductivity, as indicated by the numbers in Fig. 3c. To understand this complex LL spectrum, we adopt the model introduced by previous theoretical work[41] (see "Methods" for details). In this model, the Hamiltonian of the ABAB tetralayer graphene is unitarily transformed and decomposed into two weakly coupled effective BG systems: a light-mass sector (denoted as the b band) and a heavy-mass sector (denoted as the B band). Crucially, these effective BG bands are defined on a new set of orthonormal basis states constructed from the linear combinations of the original sublattices residing on the odd and even layers. In the presence of magnetic field, these two effective BG bands (b and B) give rise to two distinct sets of BG-like LLs. According to equation (1), the cyclotron frequency $\omega_c$ is inversely proportional to $m^*$, therefore the light-mass band b exhibits a larger energy spacing between LLs than the heavy-mass band B. The different dispersive slopes of these two sets of subbands would naturally result in a rich pattern of LL crossings. As shown in Fig. 3d, the calculated LL energy spectrum captures the widespread intersection of these two sets of LLs, where each LL is denoted by its subband (B or b) and LL index n (n+ for electron like states and n- for hole like states and $b/B_{0/1}$ for the zeroth LLs).

Guided by the calculation results, we further classify the complex experimental LLs energy spectrum into two sets of BG-like LLs based on their dispersion with magnetic field and corresponding Chern numbers (Fig 3c). Remarkably, our experimental LL energy spectrum closely matches with the theoretical results, successfully reproducing major LL crossing points, such as the crossing between $b_0$ and $B_{2+}$ at $\nu = +4$, and the crossing between $b_{2-}$ and $B_{0/1}$ at $\nu = -8/-4$. In contrast, the LL energy spectrum of ABCA-stacked tetralayer graphene exhibits no LL crossings (Supplementary Fig. 5). These results indicate that the Bernal configuration dominates the observations, likely because it covers a larger portion of the sample regardless of the local presence of moiré superlattices. Although ABCA-stacked domains are also present in a few blocks, their fraction appears too small to contribute appreciably to the measured signals, rendering their associated features undetectable in our experiment (Supplementary Fig. 4d–g).

We reveal the robustness of these QHIFM states in TDBG by temperature dependent sMIM point spectroscopy (Fig. 3e and Supplementary Fig. 6). With increasing temperature, the thermal excitation weakens the incompressible signal of these QHIFM states, ultimately causing them to vanish at critical temperatures $T_C$. We can directly extract the thermodynamic energy gap of these QHIFM states by examining the chemical potential variation across each integer filling (Supplementary Fig. 7). In Fig. 3e upper, the critical temperature $T_C$ and energy gap $\Delta$ show excellent consistent trend, with the ratio of $\Delta/k_B T_C$ generally falling between 2 - 3.5. This quantitative agreement underscores the high reliability and accuracy of our SCCM technique in probing the intrinsic energy scales of correlated quantum phases.

## LL wavefunctions, LL crossings and QHIFM phase diagram

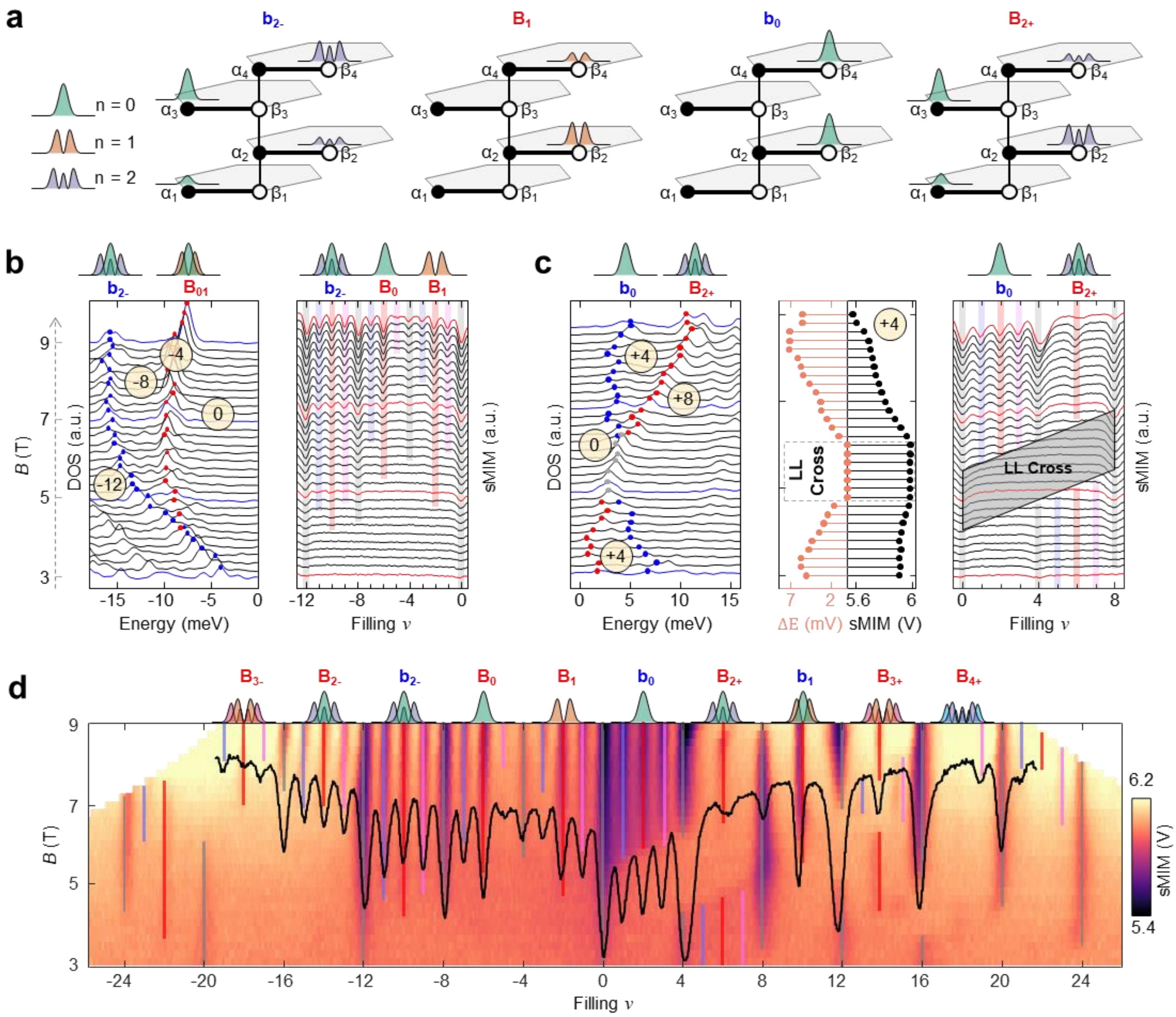

**Fig. 4 | LL wavefunctions, LL crossings and QHIFM phase diagram. a**, Schematic of single-particle LL wavefunction $b_n/B_n$ in tetralayer graphene unit cell at K valley. **b**, QHIFM ground states from $\nu$ = 0 to -12. Left: waterfall plot of energy spectrum line profile at different $B$, shows the LL crossing between $b_{2-}$ and $B_{01}$. Right: waterfall plot of sMIM as a function of filling factor $\nu$ at different $B$. The shaded bars denote the existence of broken-symmetry QH states. **c**, Left: waterfall plot of energy spectrum line profile at different $B$, shows the LL crossing between $b_0$ and $B_{2+}$. Middle: the evolution of LL energy splitting and conductivity of $\nu$ = +4 state. The consistence of energy splitting and conductivity confirms the LL crossing at $\nu$ = +4. Right: waterfall plot of sMIM as a function of filling factor $\nu$ at different $B$. A robust insulating state at $\nu$ = +1, +2, +3 manifests after LL crossing. **d**, Phase diagram of QHIFM states. Broken-symmetry QH states are identified from $\nu$-$B$ map with their LL indices and wavefunctions denoted.

Based on previous theoretical work[41], the light-mass b subband and heavy-mass B subband behave as effective BG systems. A defining feature of these effective bilayers is that their low-energy Hamiltonian contains off-diagonal terms proportional to the square of the kinetic momentum ($\pi^2$). In a perpendicular magnetic field, the momentum operators act as raising and lowering operators for the non-relativistic LL wavefunctions $\varphi_n$. Specifically, at K valley, the $n$-th LL wavefunction $\psi_n^{b/B}$ is mainly given by $\psi_n^{b/B} = (\varphi_{n-2}, \varphi_n)$ residing on the two effective basis, which are linear combinations of odd/even layer sublattices. For instance, the K-valley wavefunction of $b_{2-}$ is composed of $\varphi_2$ at $\beta_2$ and $\beta_4$ sites, and $\varphi_0$ at $\alpha_1$and $\alpha_3$ sites, and the K-valley wavefunction of $B_1$ is only $\varphi_1$ at $\beta_2$ and $\beta_4$ sites. Here $\alpha_i$ and $\beta_i$ are sublattices in the $i$th layer (Fig. 4a). The single-particle LLs wavefunction components for all low-energy subbands are summarized in Fig. 4d. Consequently, the underlying orbital character of these single-particle wavefunctions determines the electron cyclotron orbital extent, thereby governing the strength of the exchange interactions.

We show the LL crossing at $\nu = -8$ by zooming into the thermodynamic DOS deduced from chemical potential data (Fig. 4b, left). The two prominent peaks in the DOS line profile correspond to $b_{2-}$ and $B_{01}$ LLs, respectively. The QHIFM states from $\nu = 0$ to -12 originate from the crossing between $B_0$, $B_1$ and $b_{2-}$. The persistence of $\Delta_0$ and $\Delta_{-12}$ results in robust QH states at $\nu = 0$ and -12, which remain stable over a broad magnetic field range from $B$ = 3 T to 9 T. In contrast, the LL crossing event near $\nu$ = -8 triggers the emergence of the corresponding QH state above B = 4.4 T. We note that the experimental $b_{2-}$ band is a curved line in Fig. 4b, which is apparently deviated from a linear dispersion with magnetic field expected by the theoretical result (Fig. 3d). In SCCM measurement, the energy position of the $b_{2-}$ peak is not solely determined by the single-particle cyclotron energy gap described in equation (1); it is also significantly influenced by the QHIFM states between $\nu = 0$ and $\nu = -8$ driven by electron-electron interactions, leading to the observed deviation of the experimental value from the single-particle calculation. Therefore, we attribute this discrepancy to the fundamental difference between the thermodynamic chemical potential measured in our experiment and the single-particle energy band calculated in theory.

A similar LL crossing is observed at $\nu = +4$, resulting from the intersection between $b_0$ and $B_{2+}$ at $B$ = 5 T (Fig. 4c). To reveal the connection between LL crossing and the re-emergence behavior, we extract the local conductivity and energy gap of $\nu = +4$ QH state as a function of magnetic field (Fig. 4c, middle). The energy gap $\Delta_4$ is obtained by investigating the chemical potential jump at $\nu = +4$ (orange circle), whereas the conductivity is extracted by slicing through the $\nu = +4$ line (black circle) in Fig. 4c right. The LL crossing at intermediate magnetic field manifests as an enhancement in local conductivity and a simultaneous closure of the energy gap, demonstrating remarkable consistency between these two observables.

When LLs from different subbands intersect, the energy gap closes, resulting in increased conductivity and the characteristic re-emergence of QH states. Specifically, the K-valley wavefunction of $B_{2+}$ LL comprises $\varphi_2$ components at $\beta_2 + \beta_4$ sites and $\varphi_0$ components at $\alpha_1 + \alpha_3$. In contrast, the $b_0$ LL wavefunction is purely composed of $\varphi_0$ (non-relativistic zeroth LL wavefunction) localized at the $\beta_2 + \beta_4$ site (Fig. 4a). The spatially compact nature of $b_0$ wavefunction enhances the exchange interaction within the cyclotron orbital, leading to a larger correlated QHIFM energy gap than $B_{2+}$. Consequently, the enhanced Coulomb interactions in $b_0$ give rise to QHIFM states at $\nu = +5$, +6, and +7 below 5 T (Supplementary Fig. 8d), which subsequently transition to $\nu = +1$, +2, and +3 above 6 T following the LL crossing (Fig. 4c, right). Crucially, the emergence of these QHIFM states is fundamentally driven by the strong intrinsic Coulomb interactions within the LL subbands with compact wavefunctions, a mechanism that persists robustly both before and after the LL crossing. While the LL crossing defines the phase transition point, the specific ground state sequence is determined by the interaction-driven selection among these competing QHIFM states as well as single-particle effects such as the Zeeman energy and electric-field driven valley splitting. Therefore, our SCCM results clarify that the LL crossing acts as a modulator that reorganizes the sequence of the QHIFM states and shapes the QHIFM phase diagram, rather than serving as the microscopic origin of these states.

The consistency between local conductivity and the many-body LL energy spectrum establishes a direct connection between single-particle LL wavefunctions, LL crossings, and the emergence of QHIFM phases, effectively mapping out the correlated phase diagram. Building on this framework, the single-particle wavefunctions underlying these QHIFM states are detailed in Fig. 4d. Notably, these states show complete symmetry breaking on the hole side, contrasting with only valley symmetry breaking on the electron side, thereby highlighting a pronounced electron-hole asymmetry. Such valley symmetry splitting is clearly resolved by the multi-peaks feature in the zoomed-in LL energy spectrum near $\nu = +12$ (Supplementary Fig. 9), where the 4-fold degenerate $b_1$ LL splits into two distinct 2-fold degenerate sub-peaks. This splitting is consistent with a clear $\nu = +10$ QH state in corresponding local conductivity channel. As shown by the sMIM gate-dependent line scans (Fig. 2d, Supplementary Fig. 4g and Supplementary Fig. 10), the specific electron-hole asymmetry and QHIFM sequence persist robustly across different sample regions, despite the varying density and type of local charge disorder in these scans. This persistence suggests that the asymmetry is likely driven by intrinsic interaction mechanisms rather than random local potential fluctuations, although a definitive conclusion needs further theoretical and experimental investigation.

Our measurements also clearly reveal distinct behaviors of the QHIFM states in the BG and TDBG regions. The zeroth LL of BG possesses an 8-fold degeneracy (spin, valley, and orbital $N = 0,1$). In our BG data, we observe the lifting of the spin and valley degeneracy near the CNP, while the orbital degeneracy generally remains intact (Figs. 1c and 1g). In contrast, remote interlayer hopping in tetralayer graphene break this orbital degeneracy[41], creating a dense manifold of subbands at the single-particle level. As shown in Fig. 4d, these splitting-induced, multi-component LL subbands retain prominent nonrelativistic zeroth LL $\varphi_0$ and/or first LL $\varphi_1$ orbital components, including $B_{3-}$, $B_{2-}$, $b_{2-}$, $B_0$, $B_1$, $b_0$, $B_{2+}$, $b_1$ and $B_{3+}$ branches. Ultimately, this single-particle splitting, combined with the multi-component nature and wavefunction structures of the subbands, accounts for the rich QHIFM sequences uniquely observed in TDBG.

## Discussion

A high-resolution SCCM spectrum on distinct stacking domains would provide valuable insights into the origin of the QHIFM states and its relation to the moiré superlattice. However, despite our best efforts, the mechanical noise level in the dry-dilution fridge environment prevented us from resolving the moiré superlattice *in situ*. To rule out ambiguity and confirm that our spectroscopic data originates from the dominant ABAB phase, we conduct extensive sMIM line scans across large areas (> 10 μm) spanning from the adjacent BG region deep into the TDBG region (Supplementary Fig. 10 b-d). These large-scale scans reveal the ubiquitous presence of the QHIFM state sequences throughout the entire TDBG region. While the signal amplitude and the exact gate voltage onset of these states exhibit spatial fluctuation due to charge disorder, the characteristic sequence of the QH states remains robustly intact everywhere (consistent with Fig. 2d). We do not observe any regions exhibiting the distinct spectrum expected for ABCA stacking. Furthermore, we perform high-resolution sMIM 2D mapping over a 3 × 3 μm region at various filling factors (Supplementary Fig. 10e, left). Although the mechanical noise prevented direct imaging of the moiré lattice, we observe subtle spatial patterns near certain QHIFM states (e.g. $\nu = -7$ and +1), which might be the signature of the underlying moiré superlattice. Importantly, the spectroscopic signature of sMIM line scan across this area reveals a clearly resolved and uniform QHIFM sequence (Supplementary Fig. 10e, right). Despite local density variations and possible moiré-induced textures, the fact that the characteristic QHIFM sequence is observed throughout the sample confirms that the signal is dominated by the robust ABAB phase.

Moreover, our identification of the dominant ABAB phase is supported by a recent transport study on marginally TDBG[42]. They report a doping-induced sliding phase transition between the Bernal-dominant and Bernal-Rhombohedral mixed states. Their results confirm that the Bernal (ABAB) stacking is the energetically dominant ground state over large areas. While our SCCM measurements focus on the local conductivity and chemical potential rather than the global switching reported in their work, the observation of Bernal dominance is consistent across both studies. This aligns perfectly with our observation of ubiquitous QHIFM sequences and LL crossing behaviors throughout our sample.

How would moiré potential affect the QHIFM states in the ABAB domains? Previous study demonstrated marginally TDBG forms large, uniform commensurate domains by pushing the intermediate atomic stackings of the spatially evolving TDBG moiré to the domain walls and decoupling the ABCA and ABAB graphene from the effects of the superlattice potential[43]. Although a marginally TDBG moiré was used to achieve the ABCA/ABAB graphene regions, both domains are no longer affected by the much larger scale moiré, allowing the study of ABCA/ABAB graphene without any superlattice potential affecting its low-energy electronic properties. Such "moiréless" picture is further confirmed by a recent STM study[44], where the electronic structure of the relaxed domains in marginally TDBG is the same as that of intrinsic tetralayer graphene. Our sample shows similar domain configuration and twist angles. Therefore, treating the Hamiltonian as pristine tetralayer graphene is supported by the "moiréless" nature of these domains.

In summary, we establish SCCM as a versatile scanning probe microscopy technique. It enables simultaneous measurements of local conductivity and chemical potential at nanoscale, thereby allowing concurrent determination of the Chern number and thermodynamic density of states under varying magnetic field and carrier density. Our observations of robust QHIFM states, re-emergence behaviors and many-body LL energy spectrum in TDBG highlight SCCM's capacity to directly unveil the physical origins of complex quantum phases. Given its compatibility with van der Waals heterostructures and sensitivity to both electronic compressibility and conductivity, SCCM offers a broadly applicable platform for exploring correlated and topological phenomena in a wide range of two-dimensional systems.

## Methods

### Sample fabrication

The sample in this study was fabricated by employing a “cut & stack” technique developed recently with the assist of polypropylene carbonate (PPC) / polydimethylsiloxane (PDMS) stamp. A thick h-BN and bilayer graphene were exfoliated on $SiO_2$/Si substrates. The bilayer graphene was cut into two parts in advance. The h-BN flake was first picked up by PPC. Then, h-BN was used to pick up the first part of the graphene. After the transfer stage was rotated by approximately 0.2°, the second part of the graphene was picked up. In the above process, the temperature was kept at 40 $C°$. To flip the stack, the PPC film with h-BN/BG/BG was transferred to a clean PDMS, with the stack side touching with PDMS. The PPC film can be dissolved by acetone. Finally, the stack was transferred onto pre-patterned gold electrodes. Another graphite flake was used to connect the graphene and gold electrodes.

### SCCM

The SCCM experiments were carried out in a commercial cryogenic sMIM system (PrimeNano). The Pt-Ir tips from Rocky Mountain Nanotechnology LLC were used for non-contact SCCM scanning. SCCM signals were taken in a constant height mode with the scanning plane nominally ~ 30 nm above the sample surface. During SCCM measurement, the cantilever is mechanically excited by the dither piezo at its second eigenmode $f_2$=118 kHz (Supplementary Fig. 1**c**, AFM box). The phase lock loop ($PLL_2$) and $PID_2$ are used to keep cantilever oscillating at its resonance frequency and pre-setting amplitude $A_{sp}$, known as the frequency-modulated AFM. Next, a 3 GHz microwave signal is delivered to tip apex through a directional coupler and impedance matching module (Supplementary Fig.1**c**, sMIM box). The reflected microwave signal is then collected and demodulated to sMIM-Re and sMIM-Im, which characterize the electrical properties of sample. To eliminate the drifting background, the resulting oscillation amplitude of sMIM-Im with tip oscillation is demodulated at $f_2$ using $Lockin_3$ to yield d(sMIM-Im)/dz, referred as sMIM signal hereafter. Meanwhile, an ac voltage $V_{AC}$ is applied to the tip at the frequency $f_2$-$f_1$ = 100 kHz through the bias Tee (Supplementary Fig.1**c**, KPFM box). The electrical force generated at $f_1$ is given by $F_{f_1} = A_m C''(V_{DC} - V_{CPD})V_{AC}$, where $A_m$ is cantilever mechanical oscillation amplitude, the $C''$ is the second derivative of the tip-sample capacitance and $V_{CPD}$ is the contact potential difference between tip and sample. The electrical driven amplitude $A_e$ at $f_1$ = 18 kHz is demodulated by the $lockin_1$ and nullified by $PID_1$, applying a dc voltage $V_{DC} = V_{CPD}$ on tip. The chemical potential is then extracted by $V_{CPD}$.

### Extraction of thermodynamic density of states

The thermodynamic density of states, defined as $\mathrm{DOS}(\mu) = e^2 dn/d\mu$, relates to the inverse slope of the measured $\mu(n)$ dispersion. In our experiment, the linear sweep of gate voltage ensures uniform sampling intervals in carrier density $n$. Consequently, the statistical distribution of the measured $\mu$ values directly reflects the DOS structure: energy regions where $\mu(n)$ is flat accumulate a higher density of data points, corresponding to peaks in the DOS.

To reconstruct a continuous DOS spectrum from the discrete experimental data points $\{\mu_i\}$, we employed the gaussian kernel density estimation method. This approach effectively treats each data point as a gaussian wave packet and superimposes them to form the total spectrum. The estimated DOS is proportional to the sum of these gaussian kernels:

$$\mathrm{DOS}(\mu) \propto \frac{1}{Nh}\sum_{i=1}^{N}\frac{1}{\sqrt{2\pi}}\exp\left(-\frac{(\mu-\mu_i)^2}{2h^2}\right)$$

where $N$ is the total number of data points and $h$ is the width of the gaussian kernel. The width $h$ was carefully selected based on the experimental data quality to achieve an optimal balance between suppressing measurement noise and resolving fine spectral features (such as Landau level splitting). Specifically, we use a width of $h$ = 0.5 meV for the BG data and $h$ = 0.35 meV for the TDBG data. Compared to direct numerical differentiation which amplifies high-frequency experimental noise, this statistical approach yields a smooth, continuous DOS profile that allows for the unambiguous identification of intrinsic energy gaps and Landau levels.

**High-resolution moiré imaging**

An Asylum Cypher S AFM was used for high-resolution moiré imaging using sMIM and piezoresponse force microscopy (PFM). Shielded sMIM probes (ScanWave probes from PrimeNano with a spring constant of about 1 N/m) were used in contact mode for both sMIM and PFM scans. In sMIM measurement, a 3 GHz microwave excitation is directed to the sMIM probe. The reflected microwave signal is analyzed to extract the demodulated outputs as sMIM signals. The moiré signal is primarily contributed by the real part of sMIM signal (sMIM-Re). In PFM measurement, dual amplitude resonance tracking (DART) mode was used for PFM mappings to ensure strong signal-to-noise ratio. An ac bias at about 70 kHz was applied on tip to keep it oscillating at its contact resonance frequency. All the high-resolution moiré imaging measurements were conducted under ambient conditions.

**Theoretical calculation of ABAB-stacked tetralayer graphene**

The Hamiltonian of ABAB-stacked tetralayer graphene at K valley is given by[41]

$$H_{\text{ABAB}} = \begin{pmatrix} H_1 & H_2^\dagger & H_3^\dagger & 0 \\ H_2 & H_1 & H_2 & H_3'^\dagger \\ H_3 & H_2^\dagger & H_1 & H_2^\dagger \\ 0 & H_3' & H_2 & H_1 \end{pmatrix} + V_d,$$

where $H_{\text{ABAB}}$ is the $8 \times 8$ matrices expressed in the space $(\alpha_1, \beta_1, \alpha_2, \beta_2, \alpha_3, \beta_3, \alpha_4, \beta_4)$ with $\alpha_l$ and $\beta_l$ denoting the two sublattices in the $l$th layer. The block matrices are defined as:

$$H_1 = \begin{pmatrix} 0 & v_0\pi^\dagger \\ v_0\pi & 0 \end{pmatrix}, H_2 = \begin{pmatrix} -v_4\pi & \gamma_1 \\ v_3\pi^\dagger & -v_4\pi \end{pmatrix}, H_3 = \begin{pmatrix} \gamma_2/2 & 0 \\ 0 & \gamma_5/2 \end{pmatrix}, H_3' = \begin{pmatrix} \gamma_5/2 & 0 \\ 0 & \gamma_2/2 \end{pmatrix}$$

where $\pi = (k_x + ik_y)$, the velocities can be parametrized as $v_i = \sqrt{3}a_0\gamma_i/2$ for $i = 0,3,4$ with $a_0 = 0.246$ nm being the lattice constant of monolayer graphene. The diagonal term is

$$V_d = \text{Diag}[0, \delta_1, \delta_2, 0,0, \delta_2, \delta_1, 0]$$

where $\delta_1$ and $\delta_2$ represent the on-site energy differences of inequivalent sublattices. We adopt the parameter values $\gamma_0 = 3.0$ eV, $\gamma_1 = 0.44$ eV, $\gamma_2 = -0.022$ eV, $\gamma_3 = 0.2$ eV, $\gamma_4 = 0.12$ eV, $\gamma_5 = 0.01$ eV, $\delta_1 = 0.013$ eV and $\delta_2 = 0.033$ eV, which slightly differs from the parameter values available in the literature[45] and has been adjusted to capture the experimental Landau level crossing positions.

To incorporate the perpendicular magnetic field $\mathbf{B} = B\hat{z}$, the momentum $\mathbf{p} = \hbar\boldsymbol{k}$ in $H_{\text{ABAB}}$ is replaced by the kinetic momentum $\boldsymbol{\Pi} = \mathbf{p} + e\mathbf{A}$, where $\mathbf{A}$ is the vector potential satisfying $\nabla \times \mathbf{A} = \mathbf{B}$. The kinetic momentum operator has non-commuting components $[\Pi_x, \Pi_y] = -i\hbar eB$. Ladder operators are then introduced:

$$a = \frac{l_B}{\hbar\sqrt{2}}(\Pi_x - i\Pi_y), \quad a^\dagger = \frac{l_B}{\hbar\sqrt{2}}(\Pi_x + i\Pi_y)$$

where $l_B = \sqrt{\hbar/eB}$ and $[a, a^\dagger] = 1$. Using the decomposition $H_{\text{ABAB}} = M_1\pi + M_1^\dagger\pi^\dagger + M_2$, the Landau level Hamiltonian is given by:

$$H_{LL} = \frac{\sqrt{2}}{l_B}(M_1 \otimes H_a + M_1^\dagger \otimes H_a^\dagger) + M_2 \otimes I$$

where $H_{LL}$ is the $8(N+1)\times 8(N+1)$ matrix expressed in basis $|X_l,n\rangle$ and $0\leq \mathrm{n}\leq \mathrm{N}$ denotes the $n$th Landau level. $H_a$ is the $(N+1)\times(N+1)$ matrix satisfying $(H_a)_{n,n'}=\sqrt{n+1}\delta_{n+1,n'}$ and $H_a^\dagger$ is its Hermitian conjugate.

To demonstrate the decomposition into low-mass (b) and high-mass (B) sectors, we introduce a new set of orthonormal basis denoted by $|\phi_m^{(X,\rho)}\rangle$, where $X=\alpha,\beta$ labels the sublattice, $\rho=o,e$ represent odd and even layers. These basis states are defined as follows:

$$\begin{aligned}|\phi_b^{(X,o)}\rangle &= w_1|X_1\rangle - w_2|X_3\rangle, |\phi_b^{(X,e)}\rangle = w_2|X_2\rangle - w_1|X_4\rangle \\ |\phi_B^{(X,o)}\rangle &= w_2|X_1\rangle + w_1|X_3\rangle, |\phi_B^{(X,e)}\rangle = w_1|X_2\rangle + w_2|X_4\rangle\end{aligned}$$

where $X_l=\alpha_l,\beta_l$ are sublattices in $l$th layers, $w_1=\sqrt{\frac{5+\sqrt{5}}{10}}$ and $w_2=\sqrt{\frac{5-\sqrt{5}}{10}}$. The Hamiltonian $H'_{ABAB}$ in the new basis is derived by applying a unitary transformation $U$ to the original Hamiltonian $H_{\mathrm{ABAB}}$[41]. Here $U$ is formulated as:

$$U=\left(|\phi_b^{(\alpha,o)}\rangle,|\phi_b^{(\beta,o)}\rangle,|\phi_b^{(\alpha,e)}\rangle,|\phi_b^{(\beta,e)}\rangle,|\phi_B^{(\alpha,o)}\rangle,|\phi_B^{(\beta,o)}\rangle,|\phi_B^{(\alpha,e)}\rangle,|\phi_B^{(\beta,e)}\rangle\right).$$

$H'_{ABAB}$ is then given by

$$H'_{ABAB}= \quad U^\dagger H_{\mathrm{ABAB}}U=\begin{pmatrix}H_L & W\\ W^\dagger & H_H\end{pmatrix},$$

where, $H_L$ and $H_H$ are $4\times 4$ matrices for the light and heavy mass sectors, respectively. These two sectors are perturbatively coupled by the matrices $W$ and $W^\dagger$ that contain the small parameters $\gamma_2$ and $\gamma_5$.

In the presence of magnetic fields, the weights of light mass and heavy mass components in LL state $\psi$ are defined as

$$W_b=\sum_{X=\alpha,\beta}\sum_{\rho=o,e}|\langle\phi_b^{(X,\rho)}|\psi\rangle|^2, W_B=\sum_{X=\alpha,\beta}\sum_{\rho=o,e}|\langle\phi_B^{(X,\rho)}|\psi\rangle|^2,$$

where

$$|\langle \phi_m^{(X,\rho)}|\psi\rangle|^2 = \sum_n |\sum_{X',l} \langle \phi_m^{(X,\rho)}| X_l' \rangle \langle X_l', n|\psi\rangle|^2.$$

Here *n* denotes the LL index. The theoretical LL spectrum for ABAB-stacked graphene is shown in Fig. 3(d), where the weight is indicated by color. Based on these results, we further compute the density of states (DOS), as presented in Supplementary Fig. 8, where the spin and valley degrees are taken into account. Since we do not theoretically investigate interaction effects, spontaneous symmetry breaking states are not considered.

**Theoretical calculation of ABCA-stacked tetralayer graphene**

The Hamiltonian of ABCA-stacked tetralayer graphene at K valley is given by

$$H_{\mathrm{ABCA}} = \begin{pmatrix} H_1 & H_2^\dagger & H_3''^\dagger & 0 \\ H_2 & H_1 & H_2^\dagger & H_3''^\dagger \\ H_3'' & H_2 & H_1 & H_2^\dagger \\ 0 & H_3'' & H_2 & H_1 \end{pmatrix}$$

where $H_{\mathrm{ABCA}}$ is the $8 \times 8$ matrices expressed in the space $(\alpha_1, \beta_1, \alpha_2, \beta_2, \alpha_3, \beta_3, \alpha_4, \beta_4)$. The block matrices $H_1$ and $H_2$ adopt the same form as those in $H_{\mathrm{ABAB}}$, while the block matrix $H_3''$ is defined as

$$H_3'' = \begin{pmatrix} 0 & 0 \\ \gamma_2/2 & 0 \end{pmatrix}.$$

For $H_{\mathrm{ABCA}}$, we use parameters in Ref[46]: $\gamma_0 = 3.12$ eV, $\gamma_1 = 0.377$ eV, $\gamma_2 = -0.0206$ eV, $\gamma_3 = 0.29$ eV, $\gamma_4 = 0.12$ eV, $\gamma_5 = 0.025$ eV. Following the same method used in the ABAB-stacked graphene, we incorporate the perpendicular magnetic field and thereby construct the Landau level Hamiltonian for the ABCA structure. The calculated electronic band structure in the absence of magnetic field and the corresponding LL spectrum are shown in Supplementary Fig. 5.

## Data availability

The main figure data generated in this study are provided in the Source Data file. Additional data measured or analyzed during this study are available from the corresponding authors upon request. Source data are provided with this paper.

## Funding

H.-J.G. is supported by grants from the National Natural Science Foundation of China (62488201). S.Z. is supported by grants from the National Natural Science Foundation of China (12522408, 12374199, and 92463307), the National Key Research and Development Projects of China (2025YFA1212800), and the Beijing Nova Program (20240484651). F. W. was supported by National Natural Science Foundation of China (12274333 and 12550404), and National Key Research and Development Program of China (2022YFA1402400 and 2021YFA1401300). Z. C. was supported by National Natural Science Foundation of China (92477128).

## Author contributions

H.-J.G. supervised and coordinated the project. S.Z. and H.-J.G. designed the experiments. J.H., S.Z., and Y.W. conducted sMIM measurements and analyzed data. S.X. and A.G. prepared the device. B.L. and F.W. provided theoretical support. Z.C. and C.S. provided technical support and assistance. J.H., S.Z., and F.W. wrote the manuscript with inputs from all authors.

## Competing interests

The authors declare no competing interests.